\documentstyle{mn}

\input{epsf}

\newif\ifAMStwofonts %\AMStwofontstrue

\def\lesssim{\mathrel{\hbox{\rlap{\hbox{\lower4pt\hbox{$\sim$}}}\hbox{$<$}}}}

\def\gtrsim{\mathrel{\hbox{\rlap{\hbox{\lower4pt\hbox{$\sim$}}}\hbox{$>$}}}}

\def\msun{${\rm M}_{\odot}$~}

\def\ll_lsun{$\log{L/\rm L_{\odot}}$~}

\def\masa_msun{$M/ \rm M_{\odot}$~}

\def\m_mstar{$M/M_{*}$~}

\def\aap{A\&A}

\def\apj{ApJ}

\def\mnras{MNRAS}

\def\qjras{QJRAS}

\def\araa{ARA\&A}

\def\pasp{PASP}

\title[A Code for Stellar  Binary Evolution]{A Code for Stellar Binary
Evolution and its Application to the Formation of Helium White Dwarfs}

\author[O.     G.    Benvenuto,    M.    A.     De   Vito]    {O.   G.
Benvenuto\thanks{Member    of    the    Carrera    del    Investigador
Cient\'{\i}fico, Comisi\'on de  Investigaciones Cient\'{\i}ficas de la
Provincia      de      Buenos      Aires,      Argentina.       Email:
obenvenuto@fcaglp.unlp.edu.ar}, M.  A. De Vito\thanks{Fellow of
the Comisi\'on de Investigaciones  Cient\'{\i}ficas de la Provincia de
Buenos Aires, Argentina.  Email: adevito@fcaglp.unlp.edu.ar}\\
Facultad  de Ciencias  Astron\'omicas  y Geof\'{\i}sicas,  Universidad
Nacional  de  La  Plata,  Paseo  del  Bosque  S/N,  (1900)  La  Plata,
Argentina}

\date{December 17}

\pagerange{\pageref{firstpage}--\pageref{lastpage}}

\pubyear{2002}

\begin{document}

\maketitle

\label{firstpage}

\begin{abstract}

We present a numerical code intended for calculating stellar evolution
in close binary systems. In doing so, we consider that mass transfer
episodes occur when the stellar size overflows the corresponding Roche
lobe. In  such situation  we equate  the radius of  the star  with the
equivalent  radius  of the  Roche  lobe.   This  equation is handled 
implicitly together with those corresponding
to  the  whole structure  of  the star.   We  describe  in detail  the
necessary modifications to the  standard Henyey technique for treating
the  mass  loss  rate  implicitly  together with  thin  outer  layers
integrations.

We have applied  this code to the calculation of  the formation of low
mass, helium white  dwarfs in low mass close  binary systems. We found
that the  global numerical convergence properties are  fairly good. In
particular, the  onset and end  of mass transfer episodes  is computed
automatically.

\end{abstract}


%---------------------------------------------------------------------
\section{Introduction} \label{sec:intro}

Stellar evolution  in binary systems  is a classical topic  of stellar
astrophysics. The study  of stellar evolution in this  kind of systems
has great importance, since an important fraction of know stars belong
to  binary  systems.  Inasmuch  as,  if we  want  to  build a  stellar
evolution  theory  able to account  for  the  observations,  it  is
essential to study stellar evolution in binary systems.

According  to the  standard model,  the initially  most  massive star,
usually named the primary evolves faster and, as consequence of
its  nuclear   evolution,  begin  to   inflate.   If   stars  are
sufficiently close  each other  (close binary systems,  hereafter CBS)
the primary will  overflow of the ``Roche lobe''.   This lobe is found
by considering the problem of a  binary system in which the orbits are
circular, and  rotation of both  objects is synchronized  with orbital
motion.  Therefore we  have the  Largange solution  to  the restricted
three  body  problem,  considering   that  each  gas  particle  is  of
infinitesimal mass.   The Roche  lobe corresponds to  an equipotential
surface that  surrounds the stars  and represents the  limiting volume
that can be attained by the  stars of the pair before the beginning of
mass  transfer   either,  onto  the  companion,   and/or  outside  the
system. From  this moment on, the  evolution of each  star is markedly
different from the one they would have in isolation.

According to Kippenhahn \& Weigert (1967) mass transfer may begin
during core  hydrogen burning  (case A)  and also  during central
contraction after  central hydrogen  exhaustion (case  B). If  it
begins after core  helium burning it  is commonly referred  to as
case C (Lauterborn 1970). Mass transfer can occur in conservative
conditions (total mass and orbital angular momentum conserved) or
with mass and/or angular  momentum losses from the  system. These
losses can drastically affect the evolution of the star in a CBS.
For reviews on  stellar evolution in  binary systems, see,  e.g.,
Paczy{\'n}ski (1971), Iben (1985), Iben \& Livio (1993), and Taam  \&
Sandquist(2000).

In  view  of  the  large   number  of  situations  that  can   be
encountered,  it  is  not  surprising  that  CBSs  are considered as
favorable   places   for   the   occurrence   of   a  number  of
phenomenologies with high  astrophysical interest: X-ray  sources
(van der  Klis 2002),  low mass  helium white  dwarfs
(Kippenhahn, Kohl \& Weigert 1967), planetary nebulae with binary
nucleus (Iben \& Livio 1993), binary radio pulsars  (Bhattacharya
\& van den  Heuvel 1991), supernova  progenitors (Podsiadlowski,
Joss  \&  Hsu  1992).  In  this  last  case,  binary
evolution  naturally  accounts  for  the  evolution  of   the
progenitor of SN 1987A. This object was a blue supergiant, a fact
difficult to explain  in the frame  of single stellar  evolution,
but not in binary evolution (Podsiadlowski, et al. 1992).

In view of the importance of this line of research, in our Observatory
we have begun the study of CBSs. To this end, we have developed a code
to compute stellar evolution in CBSs, including
the self  consistent calculation of  the mass transfer  rate $\dot{M}$
(hereafter MTR).  The main purpose of the present paper is to describe
the numerical  techniques employed in our program  for the computation
the evolution of stars in a CBS.  This code is based on a modification
of the Henyey technique presented in Kippenhahn, Weigert \& Hofmeister
(1967) to solve  the set of difference equations  of stellar evolution
with the inclusion  of hydrodynamical effects and the  MTR. As a first
application of this new code to the case of CBSs that lead
to the formation of helium white dwarfs.

As stated  above, the  main characteristic  of our  code is that,
during mass transfer episodes, it finds the MTR in a
fully implicit way by means of a {\it single} iterative procedure. 
This is in contrast with the usual  procedure
employed  by   other  researchers   (see,  e.g.,   Podsiadlowski,
Rappaport,  \&  Pfahl  2001;  Wellstein,  Langer  \&  Braun 2001).  
In these works, in the construction of a stellar model a {\it double}
iterative  procedure  is  applied: a MTR  is estimated, and the
stellar structure is relaxed for {\it this} MTR value. Then, with
this  new  structure,  a  new  MTR  is  computed  and the stellar
structure is relaxed again. This is repeated until consistency is
achieved.  In  calculating  the  MTR  it  is  usual to employ the
physical treatment presented by Ritter (1988) which gives

\begin{equation}
\dot{M}= -\dot{M_{0}} \exp{\bigg( - \frac{R_{L}-R}{H_{P}} \bigg)};
\end{equation}

\noindent where $\dot{M_{0}}$ is the MTR for a star that exactly fills the Roche
lobe, $R_{L}$ is the equivalent  radius of the Roche lobe (see below),
$R$ is  the stellar  radius and $H_{P}$  is the  photospheric pressure
scale height.  Notice that the  double iterative loop is necessary to
get numerical  stability because  of the steep  dependence of  the MTR
upon the sizes of the star and the Roche lobe.

As the main goal of this paper is to present the numerical scheme
we  have  devised  and  not  to  produce  start  - of - the - art
evolutionary models here, for simplicity, we shall simply  impose
$R_{L} = R$. In doing so, we shall get a good global  computation
(but  not  the  details  of  the  beginning and end) of each mass
transfer episode.

The  rest   of  the  paper   is  organized  as  follows:   In  Section
\ref{sec:equa}  we  briefly  describe  the differential  equations  of
single stellar  evolution we have  to solve. In the  following Section
\ref{sec:binevol}, we describe in detail the Henyey technique and also
the code employed  to compute binary stellar  evolution. We
describe
our treatment for the orbit  of the binary, including the modification
we have considered for total mass and angular momentum losses from the
system  in \S \ref{sec:orbital}.  In  Section \ref{sec:outer}  we discuss
the critical  problem of how  to manage the outer  boundary conditions
that, in  turn, are essential in  determining the MTR  during mass
transfer episodes. Then, in  Section \ref{sec:details} we discuss some
numerical techniques  we have applied,  especially the rezoning  of the
stellar models.  In Section  \ref{sec:applic}, we present  the results
for  three different  binary systems  that  lead to  the formation  of
helium white  dwarfs.  Finally, in Section  \ref{sec:discu} we comment
on  our results, further improvements we plan to incorporate to the 
code  and the  astrophysical problems  we plan  
to study with.

%---------------------------------------------------------------------
\section{The Equations of Stellar Evolution} \label{sec:equa}

Here, we  briefly summarize the  equations of stellar evolution  to be
solved.   As   usual,  we  consider   spherically  symmetric  objects,
neglecting  rotation and  magnetic fields.   In these  conditions, the
equations of stellar structure  are (for derivation of these equations
see, e.g.,  Clayton 1968, Kippenhahn  \& Weigert 1990. For  a detailed
treatment of  hydrodynamic stellar codes, see, e.g.,  Kutter \& Sparks
1972):

\bigskip

\noindent i) the Euler equation of fluid motion

\begin{equation}
\frac{\partial  v}{\partial t}  + v  \frac{\partial v}{\partial  r}= -
\frac{1}{\rho} \frac{\partial P}{\partial r} - \frac{G\; m_r}{r^2},
\label{eq:euler}
\end{equation}

\noindent ii) the definition of velocity

\begin{equation}
\frac{\partial r}{\partial t}= v,
\label{eq:velocity}
\end{equation}

\noindent iii) the equation of mass conservation

\begin{equation}
\frac{\partial m_r}{\partial r} = 4 \pi r^{2} \rho \label{1},
\label{eq:masscons}
\end{equation}

\noindent iv) the equation of energy balance

\begin{equation}
\frac{\partial   l_r}{\partial   r}=   4   \pi   r^{2}   \rho   \left(
\varepsilon_{nuc} -  \varepsilon_{\nu} - T  \frac{\partial S}{\partial
t} \right),
\label{eq:energy}
\end{equation}

\noindent v) the equation of energy transport for the radiative case

\begin{equation}
\frac{\partial  T}{\partial  r}=   -  \frac{3  \pi}{4ac}  \frac{\kappa
\rho}{T^3} \frac{l_r}{4 \pi^2 r^2},
\label{eq:transrad}
\end{equation}

\noindent and

\noindent vi) the equation of energy transport for the convective case

\begin{equation}
\frac{\partial     T}{\partial    r}=     \nabla_{conv}    \frac{T}{P}
 \frac{\partial P}{\partial r}.
\label{eq:transconv}
\end{equation}

\noindent We employ the Schwarzschild criterium for the onset of 
convection. The symbols have their standard meaning.

%-----------------------------------------------------------------
\section{The Hydrodynamic Code for Binary Stellar Evolution}
\label{sec:binevol}

In  order  to incorporate  the  specific  phenomena  occurring in binary 
evolution we  have to make supplementary assumptions  apart from those
quoted in  the previous  section. We shall  handle the members  of our
system as  spherical objects, neglecting the  departure from spherical
symmetry  of  the  equipotentials  (e.g.,  the  Roche  lobe)  and  its
evolutionary  consequences.   Moreover,  we  shall consider  that  the
objects move along circular orbits, and that they influence each other
only through gravitational attraction (we neglect irradiation).

As  usual we  shall consider  the problem  in  Lagrangian coordinates,
considering $\xi$ the independent variable, defined as

\begin{equation}
\xi= \ln{ \left(1 - \frac{m_r}{M} \right)}. \label{eq:coord}
\end{equation}

\noindent  Radius, pressure, and  temperature are  handled by  means of
logarithmic transformations

\bigskip   \noindent   $p=\ln{P}$,\\  $\theta=\ln{T}$,\\   $x=\ln{r}$,
\bigskip

\noindent whereas $l_{r}$, $v$ are considered linearly.

For simplicity, we have written the difference equations in a centered
fashion. It means that we  have chosen to write a generic differential
equation

\begin{equation}
\frac{dy_{i}}{dx}=   F(x,y_{1},  \cdots,  y_{5});   \;\;\;\;\;\;  i=1,
\cdots, 5
\end{equation}

\noindent as a difference equation

\begin{equation}
\frac{  y_{i,j+1}-y_{i,j}}{x_{j+1}-x_{j}}  -  F(x_{j+1/2},y_{1,j+1/2},
\cdots,y_{5,j+1/2})
\end{equation}

\noindent  where $\eta_{j+1/2}=(\eta_{j+1}+\eta_{j})/2$,  being $\eta$
any quantity. The second subindex  $j$ indicates the shell of the star
for which the difference equation  is written. The results obtained in
this    way   are    completely   satisfactory    for    our   present
purposes.  However, this  would  have not  been  the case  if we  were
studying  stellar objects  suffering  shock  waves somewhere.   For
calculations of  more violent  phases we shall  have to include  non -
centered   equations    together   with   some    dissipative   effect
(e.g., artificial viscosity, see Kutter \& Sparks
1972). Temporal derivatives have been written in
the standard backwards differenced form.

In  problems with  variable  stellar mass  we  need to  be careful  at
computing temporal derivatives at constant mass. We have found it very
convenient to rewrite the derivative operator as

\begin{equation}
{\partial{     }\over{\partial{t}}}\bigg|_{m_{r}}     =     {\partial{
}\over{\partial{t}}}\bigg|_{\xi}                                      +
{\partial{\xi}\over{\partial{t}}}\bigg|_{m_{r}}\;            {\partial{
}\over{\partial{\xi}}}\bigg|_{t}.
\end{equation}

\noindent  In  this  formulation   we  get  the  dependence  of  these
derivatives with  the MTR.  This  is important because we  shall treat
the MTR as a  new variable to be relaxed in our  code.
Then  we have  to design  a Henyey  scheme for  relaxing  the internal
structure of the star, the total luminosity, the effective temperature
{\it and} (in the case of  the occurrence of mass transfer) the
MTR. 

As
usual, the calculation  of the  corrections is based  in the solution
of the lineal system

\begin{equation} {\\H} \cdot\ {\\\delta} = {\\h} \end{equation}

\noindent where ${\\H}$ is a  matrix with a structure similar to those
corresponding to the cases of  single star evolution or codes in which
the  MTR  is not  treated  self consistently.  It  has  blocks on  the
diagonal, but now, an important difference appears: now there is a non
vanishing column  due to the derivatives of  the structure
equations with respect to  the MTR. The vector ${\\\delta}$ represents
the corrections vector (see below) and ${\\h}$ is the inhomogeneity 
vector that must vanish for a correct stellar model.

Here, some words  are in  order  about the  necessity of  such a  deep
modification of the standard numerical treatments of binary evolution.
Savonije (1978) has  treated MTR as another variable  to be relaxed by
iterations but only  considered it to be relevant  in the outer layers
integration where he included most  of the change in luminosity due to
mass transfer  (See Section \ref{sec:outer} for details).   This is so
only in  the case of excluding a  very thick portion of  the star from
the grid.  In this way we are neglecting the temporal derivatives, and
so, the  inertia of  the outermost layers  of the star.   Other scheme
intended for the computation  of single and binary hydrostatic stellar
evolution has  been presented by  Ziolkowski (1970), who incorporated 
an iterative procedure  for getting the MTR. However, the
outer envelope integrations  are performed in a very  thick portion of
the  star,  and because  of  the reasons  quoted  above,  this is  not
entirely adequate for our purposes

In order  to perform  a realistic simulation  of the evolution  of the
stars, it  is quite desirable to handle  most of the mass  of the star
inside  finite - difference part of the Henyey scheme.  Notice  that  
there are  many
relevant  situations in  which the  outer  layers structure  is a  key
ingredient  in  determining the  evolution  of  the stars.   Regarding
binary evolution, we may cite for example that in order to compute the
mass transfer episodes  occurring in low mass pre  - white dwarf stars
we do need to consider  very thin outer layers integrations (see below
for more details).

Consequently, we  have to  consider an outer  layers integration  in a
portion of  the star  so thin that  a fraction,  or even most,  of the
luminosity  change (see  Section \ref{sec:outer}  for details)  due to
mass loss  during mass  transfer episodes occurs  inside the  finite -
differenced  portion of  the model.   Thus, the form of  the equation  of
energy conservation we have to consider is

\begin{equation}
\frac{   \partial  l_r   }{  \partial   m_r  }=   \varepsilon_{nuc}  -
\varepsilon_{\nu} - T \bigg( \frac{\partial S}{\partial t}\bigg|_{\xi}
+                                           {d\xi\over{dt}}\bigg|_{m_r}
{\partial{S}\over{\partial{\xi}}}\bigg|_{t} \bigg),
\label{eq:energy2}
\end{equation}

\noindent where, from Eq. (\ref{eq:coord}), we have

\begin{equation}
\frac{d\xi}{dt}\bigg|_{m_r}= \frac{  \dot{M} }{M}\; \big(  e^{-\xi} - 1
\big).
\end{equation}

\noindent Thus, {\it  we are forced} to consider  $\dot{M}$ as a extra
Henyey unknown.

Notice that, for consistency, we  should consider the total mass value
of  the star  as another  unknown to  be found  during  the relaxation
process by means of the equation

\begin{equation} M=M^{prev} + \dot{M}\; \Delta t,  \end{equation}

\noindent where $M^{prev}$ is the mass of the previous model.  Because
of the transformation of variables we have considered, the total mass
value occurs in  most of the equations of  structure. Thus, the matrix
we shall have to handle  in finding the corrections will have the
usual blocks  on the diagonal  but also a  non - vanishing  column. We
shall describe in  the rest of this section  the technique we employed
to solve the numerical problem.

The first  block of the  matrix includes the four  boundary conditions
equations that  link the  values of the  structural quantities  at the
first  meshpoint with  the effective  temperature, luminosity  and MTR
value. These  are constructed by meand of  linear interpolations among
the outer layers integrations. The resulting block has the form

$$\left[\begin{array}{ccccccc}      {\partial{h_{1}}\over{\partial{\dot
M}}}          &          {\partial{h_{1}}\over{\partial{L}}}         &
{\partial{h_{1}}\over{\partial{T_{eff}}}}                             &
{\partial{h_{1}}\over{\partial{x_{1}}}}                               &
{\partial{h_{1}}\over{\partial{l_{1}}}}                               &
{\partial{h_{1}}\over{\partial{p_{1}}}}                               &
{\partial{h_{1}}\over{\partial{\theta_{1}}}}\\
{\partial{h_{2}}\over{\partial{\dot               M}}}               &
{\partial{h_{2}}\over{\partial{L}}}                                   &
{\partial{h_{2}}\over{\partial{T_{eff}}}}                             &
{\partial{h_{2}}\over{\partial{x_{1}}}}                               &
{\partial{h_{2}}\over{\partial{l_{1}}}}                               &
{\partial{h_{2}}\over{\partial{p_{1}}}}                               &
{\partial{h_{2}}\over{\partial{\theta_{1}}}}\\ \vdots& \vdots& \vdots&
\vdots&  \vdots&  \vdots&  \vdots\\  \vdots& \vdots&  \vdots&  \vdots&
\vdots&  \vdots& \vdots\\  {\partial{h_{5}}\over{\partial{\dot  M}}} &
{\partial{h_{5}}\over{\partial{L}}}                                   &
{\partial{h_{5}}\over{\partial{T_{eff}}}}                             &
{\partial{h_{5}}\over{\partial{x_{1}}}}                               &
{\partial{h_{5}}\over{\partial{l_{1}}}}                               &
{\partial{h_{5}}\over{\partial{p_{1}}}}                               &
{\partial{h_{5}}\over{\partial{\theta_{1}}}}\\      \end{array}\right]\
\cdot\  \left[\begin{array}{c}  \delta  \dot  M\\  \delta  L\\  \delta
T_{eff}\\  \delta   x_{1}\\  \delta  l_{1}\\   \delta  p_{1}\\  \delta
\theta_{1} \end{array}\right]$$
\begin{equation}=
\left[\begin{array}{ccc} - h_{1}\\ - h_{2}\\ \vdots\\ - h_{5}
\end{array}\right],
\label{eq:15} \end{equation}

\noindent  where $h_{i},  i=1, \cdots,5$ are functions known from
linealizacion of  stellar structure equations.  Hereafter in
this section,  $L$ is $log(L/L\odot)$  and $T_{eff}$ the  logarithm of
effective temperature.

\noindent  We define  vectors  $U_{i}$, $V_{i}$,  $W_{i}$,  $X_{i}$, $i=1,
\cdots,5$, such that

\begin{equation}  \label{eq:16}  \left[\begin{array}{c}
\delta  \dot M\\  \delta L\\  \delta T_{eff}\\  \delta  x_{1}\\ \delta
l_{1}\\
\end{array}\right] =
\left[\begin{array}{cccc}  U_{1} &  V_{1} &  W_{1} &  X_{1}\\  U_{2} &
V_{2} & W_{2}  & X_{2}\\ \vdots & \vdots & \vdots  & \vdots\\ \vdots &
\vdots & \vdots & \vdots\\ U_{5} & V_{5} & W_{5} & X_{5}
\end{array}\right] \cdot\ \left[\begin{array}{c} \delta v_{1}\\ \delta
p_{1}\\ \delta \theta_{1}\\ 1
\end{array}\right].
\end{equation}

\noindent  For the  first block  are $U_{i}=0$,  $i=1, \cdots, 5$ and
$\delta v_{1}=0$.\\  From Eq.  (\ref{eq:15}) and Eq.  (\ref{eq:16}) we
can write

$$\left[\begin{array}{ccccc}  {\partial{h_{1}}\over{\partial{\dot M}}}
&                 {\partial{h_{1}}\over{\partial{L}}}                &
{\partial{h_{1}}\over{\partial{T_{eff}}}}                             &
{\partial{h_{1}}\over{\partial{x_{1}}}}                               &
{\partial{h_{1}}\over{\partial{l_{1}}}}\\
{\partial{h_{2}}\over{\partial{\dot               M}}}               &
{\partial{h_{2}}\over{\partial{L}}}                                   &
{\partial{h_{2}}\over{\partial{T_{eff}}}}                             &
{\partial{h_{2}}\over{\partial{x_{1}}}}                               &
{\partial{h_{2}}\over{\partial{l_{1}}}}\\   \vdots&   \vdots&  \vdots&
\vdots&   \vdots\\   \vdots&    \vdots&   \vdots&   \vdots&   \vdots\\
{\partial{h_{5}}\over{\partial{\dot               M}}}               &
{\partial{h_{5}}\over{\partial{L}}}                                   &
{\partial{h_{5}}\over{\partial{T_{eff}}}}                             &
{\partial{h_{5}}\over{\partial{x_{1}}}}                               &
{\partial{h_{5}}\over{\partial{l_{1}}}}\\  \end{array}\right]\  \cdot\
\left[\begin{array}{cccc}  U_{1} &  V_{1} &  W_{1} &  X_{1}\\  U_{2} &
V_{2} & W_{2}  & X_{2}\\ \vdots & \vdots & \vdots  & \vdots\\ \vdots &
\vdots & \vdots & \vdots\\ U_{5} & V_{5} & W_{5} & X_{5}
\end{array}\right]$$ \begin{equation} =\left[\begin{array}{ccc}
 - {\partial{h_{1}}\over{\partial{p_{1}}}}             &             -
{\partial{h_{1}}\over{\partial{   \theta_{1}}}}    &   -   h_{1}\\   -
{\partial{h_{2}}\over{\partial{p_{1}}}}               &              -
{\partial{h_{2}}\over{\partial{  \theta_{1}}}} &  -  h_{2}\\ \vdots  &
\vdots    &    \vdots\\    \vdots    &    \vdots    &    \vdots\\    -
{\partial{h_{5}}\over{\partial{p_{1}}}}               &              -
{\partial{h_{5}}\over{\partial{ \theta_{1}}}} & - h_{5}\\
\end{array}\right], \end{equation}

\bigskip

\noindent  and from  this  we can  obtain  vector components  $V_{i}$,
$W_{i}$, $X_{i}$, $i=1, \cdots,5$. In  a similar way we can generalize
this mechanism for any interior block.  If we propose

\begin{equation} \delta  \dot M  = A_{n} \delta  v_{n} +  B_{n} \delta
p_{n} + C_{n} \delta \theta_{n} + D_{n}, \end{equation}

\noindent we find, generalizing the equations of the interiors blocks,

\begin{equation} \begin{tabular}{lclclclcl}  $A_{n} = A_{n-1} U_{5n-4}
+  B_{n-1} U_{5n-3}  + C_{n-1}  U_{5n-2}$\\ $B_{n}=A_{n-1}  V_{5n-4} +
B_{n-1}  V_{5n-3}  +  C_{n-1}  V_{5n-2}$\\ $C_{n}=A_{n-1}  W_{5n-4}  +
B_{n-1}  W_{5n-3}  +  C_{n-1}  W_{5n-2}$\\ $D_{n}=A_{n-1}  X_{5n-4}  +
B_{n-1} X_{5n-3} + C_{n-1} X_{5n-2} + D_{n-1}$.  \end{tabular}
\end{equation}

\noindent The equation  that leads to intermediates blocks  are in the
form
$$\left(\delta  v_{n} \alpha_{i}  +  \delta p_{n}  \beta_{i} +  \delta
\theta_{n}       \gamma_{i}\right)        +       \delta       x_{n+1}
{\partial{g_{i}}\over{\partial{x_{n+1}}}}     +     \delta     l_{n+1}
{\partial{g_{i}}\over{\partial{l_{n+1}}}} =$$
\begin{equation}       -\delta         v_{n+1}
{\partial{g_{i}}\over{\partial{v_{n+1}}}}        -\delta       p_{n+1}
{\partial{g_{i}}\over{\partial{p_{n+1}}}}     -\delta     \theta_{n+1}
{\partial{g_{i}}\over{\partial{\theta_{n+1}}}} - \xi_{i},
\label{eq:20}
\end{equation} $i=1, 2, \cdots, 5\\ n=2, 3, \cdots, N-2$.\\

\noindent The $g_{i}$ functions are know from linealizacion of stellar
structure  equations  in  the  intermediate shells,  and  $\alpha_{i},
\beta_{i}, \gamma_{i}, \xi_{i}$ are defined as \bigskip

\begin{tabular}{lclclclcl}  $\alpha_{i}$  =  $A_{n}\
{\partial{g_{i}}\over{\partial{\dot       M}}}$      +      $U_{5n-1}\
{\partial{g_{i}}\over{\partial{x_{n}}}}$           +          $U_{5n}\
{\partial{g_{i}}\over{\partial{l_{n}}}}$                              +
${\partial{g_{i}}\over{\partial{v_{n}}}},$\\ $$ & $$ & $$ & $$ & $$ &
$$    &     $$    &    $$    &    $$\\     $\beta_{i}$    =    $B_{n}\
{\partial{g_{i}}\over{\partial{\dot       M}}}$      +      $V_{5n-1}\
{\partial{g_{i}}\over{\partial{x_{n}}}}$           +          $V_{5n}\
{\partial{g_{i}}\over{\partial{l_{n}}}}$                              +
${\partial{g_{i}}\over{\partial{p_{n}}}},$\\ $$ & $$ & $$ & $$ & $$ &
$$    &    $$    &     $$    &    $$\\    $\gamma_{i}$    =    $C_{n}\
{\partial{g_{i}}\over{\partial{\dot       M}}}$      +      $W_{5n-1}\
{\partial{g_{i}}\over{\partial{x_{n}}}}$           +          $W_{5n}\
{\partial{g_{i}}\over{\partial{l_{n}}}}$                              +
${\partial{g_{i}}\over{\partial{\theta_{n}}}},$\\ $$ & $$ & $$ & $$ &
$$   &    $$   &    $$   &   $$    &   $$\\   $\xi_{i}$    =   $D_{n}\
{\partial{g_{i}}\over{\partial{\dot       M}}}$      +      $X_{5n-1}\
{\partial{g_{i}}\over{\partial{x_{n}}}}$           +          $X_{5n}\
{\partial{g_{i}}\over{\partial{l_{n}}}}$ + $1.$\\ $$ & $$ & $$ & $$ &
$$ & $$ & $$ & $$ & $$\\ \end{tabular}

\noindent Written in matrix form Eq. (\ref{eq:20}), with

$$\left[\begin{array}{c}   \delta   v_{n}\\   \delta  p_{n}\\   \delta
\theta_{n}\\ \delta x_{n+1}\\ \delta l_{n+1}\\
\end{array}\right] =  \left[\begin{array}{cccc} U_{5n+1} &  V_{5n+1} &
W_{5n+1}  & X_{5n+1}\\  U_{5n+2} &  V_{5n+2} &  W_{5n+2}  & X_{5n+2}\\
\vdots  & \vdots  &  \vdots &  \vdots\\  \vdots &  \vdots  & \vdots  &
\vdots\\ U_{5n+5} & V_{5n+5}  & W_{5n+5} & X_{5n+5} \end{array}\right]
\cdot\  $$  \begin{equation}  \left[\begin{array}{c} \delta  v_{n+1}\\
\delta p_{n+1}\\ \delta \theta_{n+1}\\ 1 \end{array}\right]
\label{eq:21}
\end{equation}

\noindent and  substituting Eq.  (\ref{eq:21})  in matrix form  of Eq.
(\ref{eq:20}),  we  obtain an  equation  that  allow  us to  find  the
components for vectors $U, V, W, X$:

$${\left[\begin{array}{ccccc}  \alpha_{1} &  \beta_{1} &  \gamma_{1} &
{\partial{g_{1}}\over{\partial{x_{n+1}}}}                             &
{\partial{g_{1}}\over{\partial{l_{n+1}}}}\\  \alpha_{2} &  \beta_{2} &
\gamma_{2}      &      {\partial{g_{2}}\over{\partial{x_{n+1}}}}     &
{\partial{g_{2}}\over{\partial{l_{n+1}}}}\\ \vdots & \vdots & \vdots &
\vdots  &  \vdots\\ \vdots  &  \vdots &  \vdots  &  \vdots &  \vdots\\
\alpha_{5}        &       \beta_{5}        &        \gamma_{5}       &
{\partial{g_{5}}\over{\partial{x_{n+1}}}}                             &
{\partial{g_{5}}\over{\partial{l_{n+1}}}}\\ \end{array}\right]} \cdot\
$$  $$\left[\begin{array}{cccc}  U_{5n+1}  &  V_{5n+1}  &  W_{5n+1}  &
X_{5n+1}\\ U_{5n+2} & V_{5n+2} & W_{5n+2} & X_{5n+2}\\ \vdots & \vdots
& \vdots & \vdots  \\ \vdots & \vdots & \vdots &  \vdots \\ U_{5n+5} &
V_{5n+5} & W_{5n+5} & X_{5n+5}\\ \end{array}\right] = $$

\begin{equation}
 \left[\begin{array}{cccc} - {\partial{g_{1}}\over{\partial{v_{n+1}}}}
&      -       {\partial{g_{1}}\over{\partial{p_{n+1}}}}      &      -
{\partial{g_{1}}\over{\partial{\theta_{n+1}}}}   &   -   \xi_{1}\\   -
{\partial{g_{2}}\over{\partial{v_{n+1}}}}              &             -
{\partial{g_{2}}\over{\partial{p_{n+1}}}}              &             -
{\partial{g_{2}}\over{\partial{\theta_{n+1}}}} &  - \xi_{2}\\ \vdots &
\vdots &  \vdots & \vdots \\  \vdots & \vdots  & \vdots & \vdots  \\ -
{\partial{g_{5}}\over{\partial{v_{n+1}}}}              &             -
{\partial{g_{5}}\over{\partial{p_{n+1}}}}              &             -
{\partial{g_{5}}\over{\partial{\theta_{n+1}}}} & - \xi_{2}\\
\end{array}\right].$$
\end{equation}

\noindent In  the last block  we have $\delta  l_{n} = \delta  x_{n} =
\delta v_{n} =0$, thus, rewriting  the previous matrix equation in the
usual way and proposing, like we have done in this derivation

\begin{equation}  %\label{14}
\left[\begin{array}{c}   \delta  v_{N-1}\\  \delta   p_{N-1}\\  \delta
\theta_{N-1}\\ \end{array}\right]  = \left[\begin{array}{ccc} V_{5N-4}
& W_{5N-4}  & X_{5N-4}\\ V_{5N-3}  & W_{5N-3} & X_{5N-3}\\  V_{5N-2} &
W_{5N-2} & X_{5N-2}
\end{array}\right] \cdot\ \left[\begin{array}{c} \delta p_{N}\\ \delta
\theta_{N}\\ 1 \end{array}\right] \end{equation}

\noindent we can  compute all components of vectors $U,  V, W, X$, and
then  all the  corrections.  Finally,  at  the surface  we have  total
luminosity, effective temperature and the MTR for each model.

As stated above,  in this work we assume that  mass transfer occurs if
stellar  radius $R$ overflows the  effective radius  $R_{L}$ of  the Roche
lobe, calculated with the formula of Eggleton (1983),

\begin{equation}
\frac{R_{L}}{a}= \frac  { 0.49 \cdot q^{2/3}  } { 0.6  \cdot q^{2/3} +
                       \ln{ (1 + q^{1/3}) } }
\label{eq:eggleton}
\end{equation}

\noindent where $a$ is the orbital separation and $q \equiv M / M_{2}$
the mass ratio of the binary components.

When  a star losses  mass, either  by a  stellar wind  in the  case of
isolated star or  by transfer in CBS evolution,  not only the external
layers, but all stellar structure  is affected by this phenomenon.  An
outer layer that was internal,  after a certain time interval, it will
be in the stellar surface. Therefore,  we need to consider in the code
this  important effect.  We  have  found it  enough  to simulate  this
outflow by means of a  linear interpolation in the abundances per mass
fractions.   Then  we  should  check the  meshpoint  distribution  for
consistency.

It is  important to stress here  that in doing so,  we are introducing
noise in the  chemical profile of the donor  star. Presumably, this is
responsible  (at least  in part)  for the  noise in  the MTR  vs. time
relations shown below for the  specific systems we have computed. This
is especially  noticeable in the  case of donor stars  with convective
envelopes. Evidently, a  better strategy should be to  employ a moving
grid tailored to follow the  outwards motion of the stellar layers. We
plan to incorporate this refinement in the near future.

%--------------------------------------------------------------------------
\section{The Orbital Evolution} \label{sec:orbital}

As a first approximation, we can perform conservative binary evolution
calculations. In this  case we consider  total mass and
orbital angular moment  as constants.   However, we  expect the
occurrence  of total  mass  and  angular momentum  losses  to be  very
important in some astrophysical  situations of interest. Thus, we have
included in our code, these phenomena.

From  the  definition  of  total  angular orbital  moment,  and  using
Kepler's third law, we can write

\begin{equation} \label{eq:adot}
{{\dot     a}\over    {a}}=2{{\dot     J}\over{J}}     +    \dot     M
\left({{1}\over{M+M_{2}}}   -   {{2}\over{M}}\right)   +  \dot   M_{2}
\left({{1}\over{M+M_{2}}} - {{2}\over{M_{2}}}\right)
\end{equation}

\noindent  where  $\dot{J}/J$   represents  energetic  losses  due  to
different processes, $M$ and $M_{2}$  the masses, and $\dot M,
\dot M_{2}$ the MTRs of the lossing - accreting stars.

If  we consider  non  -  conservative mass  transfer,  this causes  an
angular  loss from  the system.   We follow  the method  developed for
Rappaport,  Joss \&  Webbink  (1982) and  Rappaport,  Joss \&  Verbunt
(1983). It is specified by  two free parameters, the fraction $\beta$,
of  the mass lost  by the  primary star that  is accreted  by the
secondary  star $2$ and  the specific  angular momentum,  $\alpha$, of
matter lost from the system in units of $2 \pi a^{2}/P$, so that

\begin{equation} \delta J= \alpha  \delta M (1-\beta) {{2 \pi
a^{2}}\over{P}} \end{equation}

\noindent where  $\delta M$ is an  incremental mass lost  by the donor
star,  $\delta  J$ the  incremental  angular  momentum  of the  matter
lost by the system, and $a$  and $P$ are the  orbital semi major
axis and  period, respectively.   We assume that  the orbit  is always
well approximated by  a circle of radius $a$ (where  $a$ is a function
of  time).  If we  consider only  angular momentum  losses due  to the
ejection of matter from the system, the last equation can be rewritten
as the differential equation

\begin{equation}  \label{eq:deltaj} \delta  J =  \alpha (1  -  \beta) \sqrt{G
\left(M+M_{2}\right) a} \ \ \delta M. \end{equation}

Using  Kepler's third  law  to  eliminate $P$,  and  combined with  an
expression for the total systemic angular momentum,

\begin{equation} J = \sqrt {{{G a} \over {M + M_{2}}}} M M_{2}.
\end{equation}

\noindent  Eq. (\ref{eq:deltaj})  constitutes a  differential equation
for  $J$  (or  $a$) as  a  function  of  $M$  (we have  neglected  the
rotational angular  momentum of the  components stars).   Then,  we 
write  the loss  of angular momentum for the matter ejection as

\begin{equation} {{d  ln J_{ME}}  \over {dt}} = {{\alpha (1  - \beta)
\sqrt{ \left(M+M_{2}\right)Ga} } \over {J}} \ \ \dot M.
\label{eq:jdot1} \end{equation}

\noindent Angular  momentum  loss  due  gravitational  radiation  is  
calculated according to the standard formula (Landau \& Lifshitz 1959)

\begin{equation} \label{eq:jdot2} \frac{d\ln{J_{GR}}}{dt}=  - {{32 G^{3}
\mu}\over{5 c^{5}}} {{M_{tot}^{2}}\over{a^{4}}} \end{equation}

\noindent where $M_{tot}=M + M_{2}$ and $\mu = M M_{2} / (M + M_{2})$,
$G$ and $c$ are the  gravitational constant and vacuum speed of light,
respectively.

To calculate the angular momentum loss due to magnetic braking, we use
the prescription of Rappaport, Joss  \& Verbunt (1983), which is based
on the magnetic - braking law of Verbunt \& Zwaan (1981),

\begin{equation} {{dJ_{MB}} \over{dt}} = - 3.8 \times 10 ^{-30} M R^{4}
\omega^{3} dyn\; cm. \label{eq:jdot3} \end{equation}

\noindent  where $\omega$  is the  angular rotation  frequency  of the
donor star,  assumed to  be synchronized whit  the orbit.   We include
full magnetic braking when the star has a sizeable convective envelope
embracing   a   mass  fraction   $\geq   0.02$.    If  we   substitute
Eqs.  (\ref{eq:jdot1})  -  (\ref{eq:jdot3})  in  (\ref{eq:adot}),  and
consider in this last equation

\begin{equation}
\dot M_{2} = - \beta \dot M,
\end{equation}

\noindent  in  view  of  the   definition  of  $\beta$,  we  obtain  a
differential  equation  for  the  orbital  separation,  which  has  no
analytical solution.

Let us remark  that the semiaxis of  the orbit is dependent  upon the MTR
and so,  the equivalent radius of  the Roche lobe, which  in turn, has
been assumed to be the radius  of the star. Thus, during mass transfer
episodes,  to  be consistent  with  the  iterated  value of  MTR, 
we  need to  perform an  orbital integration  for each
iteration.   As  the equations  for  the  orbital  evolution are  well
behaved,  we decided  to handle  them with  a standard  Runge  - Kutta
technique (Press,  et al.  1992). This  is very fast  and represents a
tiny increment of the total numerical effort.

%-------------------------------------------------------------------------
\section{Handling Outer Layers Integrations} \label{sec:outer}

Here  we  shall  treat  the  problem  of  handling  the  outer  layers
integrations  adequately   for  the  case  in  which   there  is  mass
transfer. If not, outer layers are managed as in
Kippenhahn,  et al. (1967).   We refer  the reader  to that  paper for
further details  in this simple  situation and only describe  here its
modifications to account for the occurrence of mass transfer.

In  the  outer  layers,   we  integrate  the  equations  of  structure
neglecting  temporal  derivatives  and  the velocity  but  taking  the
acceleration of these  layers into account. Then, we  have a system of
ordinary differential  equations. In particular, the  equation for the
luminosity we consider  is (see Eq \ref{eq:energy2})\footnote{However,
we  warn  the  reader  that  other  authors  have  employed  different
procedures. See especially, Podsiadlowski, et al. (2001).}

\begin{equation}
\frac{d   l_r   }{   d   m_r  }=   -   T   {d\xi\over{dt}}\bigg|_{m_r}
{\partial{S}\over{\partial{\xi}}}\bigg|_{t}.
\end{equation}

\noindent In the  case of  constant mass evolution,  it simply implies  that the
luminosity is  constant in the outer  layers. However, in  the case of
mass transfer/loss episodes, it  modifies the profile of luminosity in
a sizeable  way. For example,  for radiative envelopes, which  have an
entropy  increase  outwards, this  equation  predicts  a  drop in  the
luminosity.  This  is an  important effect to  be considered  (see the
previous sections).  If  we take a little amount of  mass $M_{OLI}$ in 
the outer
layers integrations (e.g. $M_{OLI}/M  \leq 10^{-6}$), it is present in
the outer layers  integrations as well as in  the finite - differenced
portion of the star.

In order to  perform an adequate treatment of the  outer layers of the
mass losing component  of the pair we need to  be careful.  Because of
the way we  have planned our iterative scheme, we  need to compute the
boundary condition  equations and its derivatives with  respect to the
values  of  the  dependent  variables  at  the  first  meshpoint,  the
luminosity and effective temperature of the star and also with respect
to the MTR $\dot{M}$.

In view of the need for  values of the derivatives with respect to the
MTR,  we have  found it  very convenient  to generalize  the triangles
method presented in  Kippenhahn, et al.  (1967) in  the following way.
Let us consider a MTR $\dot  M^{(1)}$ and a timestep $\Delta t$; then,
the mass of the star will  be $M^{(1)}= M^{prev} + \dot M^{(1)} \Delta
t$. Now,  we construct  a triangle  in the HR  diagram as  proposed by
Kippenhahn, et al.  (1967), i.e., we perform three integrations with

\bigskip

\noindent
i)  $\log{{\tt   L}/L_{\sun}},  \log{{\tt  T}_{eff}},   M^{(1)},  \dot
M^{(1)}$,\\  ii)  $\log{{\tt  L}/L_{\sun}}  +  \Delta\log{L/L_{\sun}},
\log{{\tt  T}_{eff}},   M^{(1)},  \dot  M^{(1)}$,\\   iii)  $\log{{\tt
L}/L_{\sun}},  \log{{\tt T}_{eff}}+\Delta\log{T_{eff}},  M^{(1)}, \dot
M^{(1)}$,

\bigskip   \noindent   where   $\log{{\tt  L}/L_{\sun}}$,   $\log{{\tt
T}_{eff}}$ corresponds to one vertex of the triangle in the HR diagram
and $\Delta\log{L/L_{\sun}}$,  $\Delta\log{T_{eff}}$ are fixed values.
We  also need  to perform  integrations with  another MTR  value $\dot
M^{(2)}=\dot M^{(1)} + \Delta \dot{M}$ so that, for the same timestep,
the mass of  the star is now $M^{(2)}= M^{prev}  + \dot M^{(2)} \Delta
t$ and the triangle in the HR diagram is

\bigskip

\noindent
iv)  $\log{{\tt  L}/L_{\sun}},   \log{{\tt  T}_{eff}},  M^{(2)},  \dot
M^{(2)}$,\\  v)   $\log{{\tt  L}/L_{\sun}}  +  \Delta\log{L/L_{\sun}},
\log{{\tt   T}_{eff}},  M^{(2)},   \dot  M^{(2)}$,\\   vi)  $\log{{\tt
L}/L_{\sun}}, \log{T_{eff}} + \Delta\log{{\tt T}_{eff}}, M^{(2)}, \dot
M^{(2)}$.

\bigskip \noindent As the model is iterated, $\dot{M}$  changes. We require the MTR to be
in  between the  values at  which the  envelopes were  computed, i.e.,
$\dot M^{(1)} \leq \dot{M} \leq  \dot M^{(2)}$.  If not, we change the
values  of  $\dot  M^{(1)}$,  $\dot M^{(2)}$.   Also,  if  $(\log{{\tt
L}/L_{\sun}},  \log{{\tt  T}_{eff}})$,  were  outside the  assumed  HR
triangle we change  it with the algorithm presented  in Kippenhahn, et
al. (1967) but now with the same mass and MTR values.

Now, we perform a linear  interpolation in $\dot{M}$ for the values of
the dependent variables at the bottom of the envelope.  Notice that in
this way we  are automatically interpolating for the  correct value of
the  mass  of the  star.   Also, we  compute  the  derivatives of  the
dependent  variables with respect  to the  MTR at  each vertex  of the
triangle. Then, all the relevant quantities are found by means of a linear
bidimensional  interpolation  inside the  triangle  in  the  HR as  in
Kippenhahn, et al. (1967).

The  experience we have  gained with  our code  indicates that  a very
convenient transformation for the MTR variable is

\begin{equation}  \sigma  \equiv  \ln{\bigg(  -  \frac{\dot{M}}{1
M_{\sun}/y} \bigg)}. \end{equation}

\noindent Notice that,  in this way,  the MTR has  its correct sign  every time.
The envelopes are computed  for given values $\sigma_1$ and $\sigma_2=
\sigma_1 + \Delta\sigma$. For most purposes we have found it enough to
set   $\Delta\sigma  \sim   0.5$,  $\Delta\log{L/L_{\sun}}=   0.02$, and
$\Delta\log{T_{eff}}=0.005$.    This  represents  a   good  compromise
between the amount of required envelope integrations and the precision
in the  interpolations and allows  the whole iteration to  converge in
few steps.

We compute  a set of six  outer envelope integration in  the case that
the  chemical composition  of the  outermost portion  of the  star has
changed.   Otherwise, if  the values  of  $\sigma$, $\log{L/L_{\sun}}$
$\log{T_{eff}}$  fall outside  the  prism corresponding  to the  outer
boundary integrations  employed for the previous model,  we change the
necessary vertex in order to minimize the  number of outer
integrations. This is important because they are time consuming.

Let  us describe  how  to compute  the  moment of  the  onset of  mass
transfer.   We assume that  we have  begun our  sequence of  models in
conditions  at  which the  radius  of the  star  is  smaller than  the
corresponding to  the sphere  equivalent to the  Roche lobe,  given by
Eq. (\ref{eq:eggleton}).  At each moment  we compute the  structure of
the  star  together  with  the  orbital  evolution.  Notice  that  the
computation  of the  orbit  is necessary  in  view that  there may  be
angular  momentum dissipation mechanisms  operating even  without mass
transfer  (e.g.  gravitational  radiation). If  the components  of the
pair are  close enough each  other, at some  moment the radius  of the
most  massive  component of  the  pair\footnote{Notice, however,  that
there are  situations in  which, e.g., a  canonical 1.4  \msun neutron
star has,  e.g., a lower mass  main sequence companion.  In this case,
obviously,  the fastest  evolving object  will be  the low  mass one.}
will  overflow  the  Roche  radius.   At this  stage  we  discard  the
evolutionary model.  Then, we  guess  a MTR  low  enough  and try  to
converge the full algorithm presented  above. If the MTR value guessed
is too  high and/or the  time step is  too long we shall  exchange too
much mass.  This usually makes the iteration divergent, but even if it
converges it  should be discarded.   Then, we halve the  timestep, and
perform  a constant  mass integration  sequence up  to Roche
lobe overflow.  Usually  this procedure  is enough to  get a  plausible initial
mass lossing  model. From  then on, the code works  in a  very similar
fashion  compared to  the  case of  constant  mass sequences.   This
procedure for  getting the  structure of the  first model in  the mass
transfer  stage  is  certainly   more  straightforward  than  the  one
described by Ziolkowski (1970).

A more  difficult point is to find  a self consistent end  of the mass
transfer stage. Usually, we set  a low MTR threshold, much smaller than
the value we guessed for the  beginning of the mass transfer stage. As
the star  gets near the end of  the mass exchange epoch,  the MTR goes
down,  sometimes  very  suddenly.   Meanwhile  the MTR  is  above  the
referred  threshold, we  assume  that  the next  model  will be  still
loosing mass,  otherwise we perform a constant  mass evolutionary step
and then  we compare  the radius of  the star  with that of  the Roche
lobe. If the  star is smaller, we assume that  the mass exchange stage
has finished  and from there on,  we continue the  evolutionary calculation
with  constant   mass.   Otherwise,  we  discard   the  mass  constant
integration and  perform a mass  exchanging integration with  a halved
timestep. We have found it to work in most situations.

%---------------------------------------------------------------------
\section{Other Relevant Numerical Details} \label{sec:details}

Another critical,  and certainly non trivial problem  is the rezoning.
This   is   unavoidably   necessary   during   the   course   of   the
evolution. Here, rezoning is more  problematic compared to the case of
the  evolution of  isolated  stars. Notice  that,  the stellar  radius
represents  a global characteristic  of a  star that,  in the  case of
isolated  stars it  appears only  in the  boundary  condition relating
effective temperature with luminosity.  However, in the case of binary
evolution  during  mass transfer  episodes,  it  also  appears in  the
condition that equates the radius of star with the radius of the Roche
lobe.  This equation, in turn,  largely determines the MTR, a quantity
that deeply affects the structure of the whole star.

At rezoning we unavoidably  introduce numerical noise that affects the
radius of the star. Thus, it  is not surprising that, in some critical
stages of mass transfer, rezoning  spoils the convergence of the code.
In most of  these cases we recovered convergence  simply performing no
rezoning.  Even in well behaved cases,  the noise in the radius of the
star produces  large oscillations  in the value  of the MTR.   This is
especially true in the case of donor stars with convective envelopes.

In  the present  version of  the code  we have  simply  considered the
rezoning criteria  described in Kippenhahn,  et al. (1967):  We simply
test if the variation of  the independent and dependent variables in a
zone is larger  than a given threshold, if so, we  add a new meshpoint
dividing the  old zone in  two.  At the  new meshpoint we  compute the
value of the functions by means of linear interpolations.  Conversely,
if the variation of all variables are below another prefixed value, we
remove a  meshpoint merging  two zones  in one.  As  it will  be shown
below, this procedure reveals as adequate for the purpose of computing
the  evolution of  low mass  CBS that  give rise  to the  formation of
helium white dwarfs and also for the case of high mass CBSs.
However, in some cases, we have  been not able to follow the evolution
in very fast  stages, like a born again phenomenon  of a $\approx 0.8$
\msun star formed from a 5 \msun  object in a 3 days period system. We
presume such problems to be due to the rezoning.

Notice that the new meshpoints introduced by linear interpolations are
not  solution  of  the  structure  equations and  participate  in  the
calculation  in  the temporal  derivatives  of pressure,  temperature,
radius and  velocity. We expect  to improve significantly  the present
rezoning scheme by introducing  a Hermite interpolation for the values
of the dependent variables at inserting new meshpoints as suggested by
Wagenhuber \&  Weiss (1994).  Hermite  interpolation in a zone  can be
performed employing a cubic expression whose coefficients are computed
using the values of the function and their derivatives at the edges of
the zone.  (see  Wagenhuber \& Weiss 1994 for  more details).  We plan
to perform these changes in the near future.

%---------------------------------------------------------------------
\section{Application  of the  Code  to  Some Low  Mass Close
Binary Systems} \label{sec:applic}

Before describing the calculations we  have performed as a test of our
new  binary  evolutionary  code,  we  have to  describe  the  physical
ingredients  we have  included  in the  present  version.  Because  of
convenience in  the process of preparing  the code, we  have chosen to
include  some very simplified ingredients, mainly  in the  nuclear description  of the
problem.   We have  included the  equation of  state,  OPAL opacities,
conductive and molecular opacities, and neutrino emission processes as
in Althaus,  Serenelli \&  Benvenuto (2001). We  have employed  a very
simple  nuclear   reaction  network   in  which  we   have  considered
equilibrium  expressions for proton  - proton,  CNO cycles  and helium
burning cycles  given in Clayton (1968).  We  have followed explicitly
the abundances of $^1H$,  $^4He$, $^{12}C$, and $^{16}O$ employing the
standard Arnett  \& Truran (1969) implicit matrix  algorithm.  We have
neglected diffusion processes.

In order  to test our numerical  scheme we have chosen  to compute the
evolution of  three low  mass, CBSs that  undergo non  - conservative,
Class  A and  B mass  transfer episodes  from the  main sequence  to a
highly evolved white dwarf configuration.

In all cases we have assumed  $\alpha= 0.5$ and $\beta= 0$, i.e.  that
the mass lost by the primary star leaves the system (no accretion onto
the secondary) and half of  the specific angular momentum is gone away
carried by such material.  In particular we have considered \\

\noindent
1) $M= 2.0$  \msun, $q_0= 1.5$, $P_0=  1.5$ days \\ 2)  $M= 2.0$ \msun,
$q_0= 1.5$,  $P_0= 0.7$ days  \\ 3) $M=  1.4$ \msun, $q_0=  1.0$, $P_0=
1.0$ days. \\

\begin{figure*}
\epsfysize=550pt \epsfbox{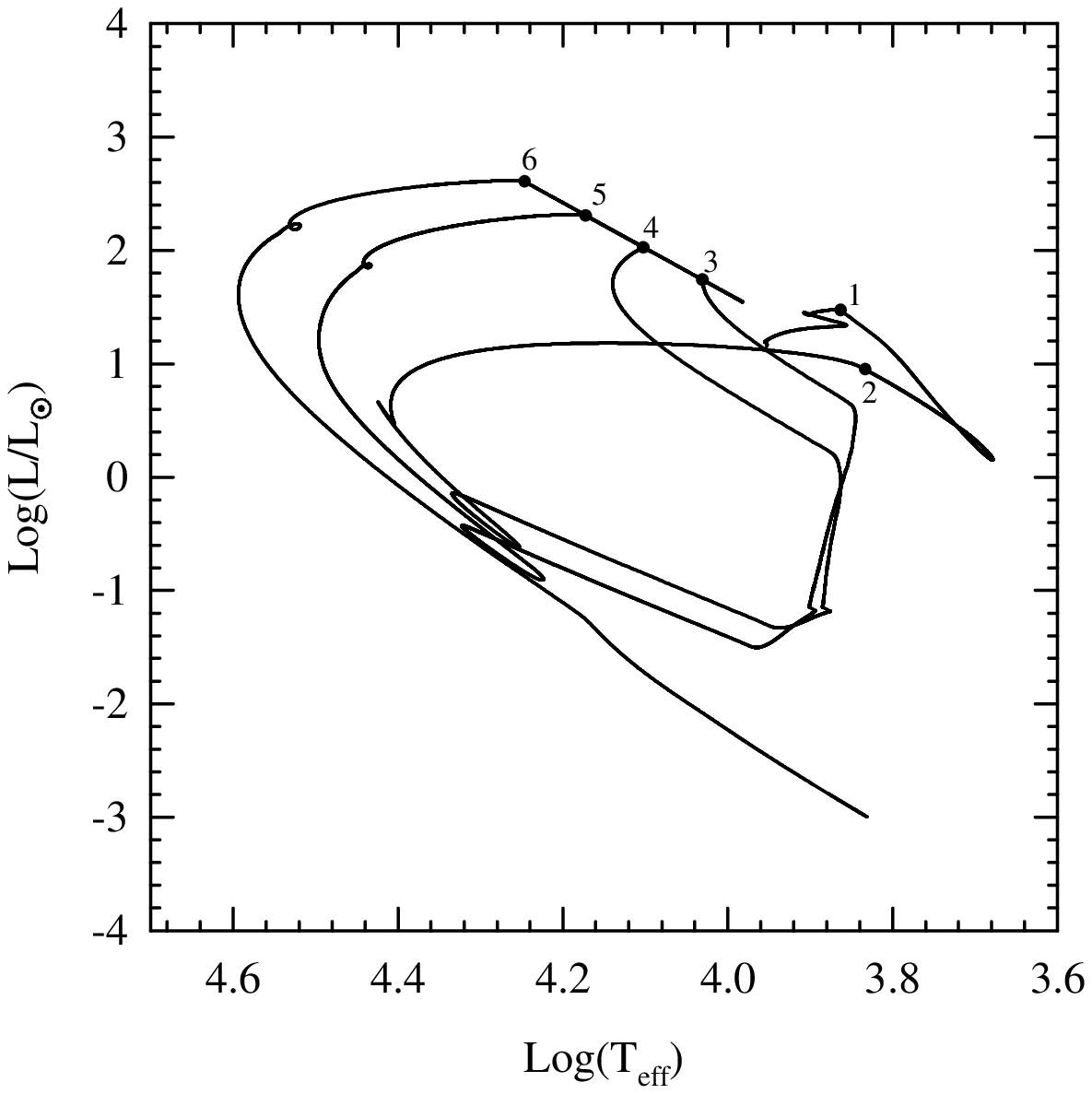}
\caption{The evolutionary track for  the primary component of system 1
($M= 2.0$  \msun, $q_0=  1.5$, $P_0= 1.5$  days). Dots labeled  with odd
(even)   numbers   indicate  the   onset   (end)   of  mass   transfer
episodes. Notice  that the  star undergoes two  hydrogen thermonuclear
flashes  each of  them  producing  the large  loops  shown.  For  more
details, see text.} \label{fig_HR_1}
\end{figure*}

\begin{figure*}
\epsfysize=550pt \epsfbox{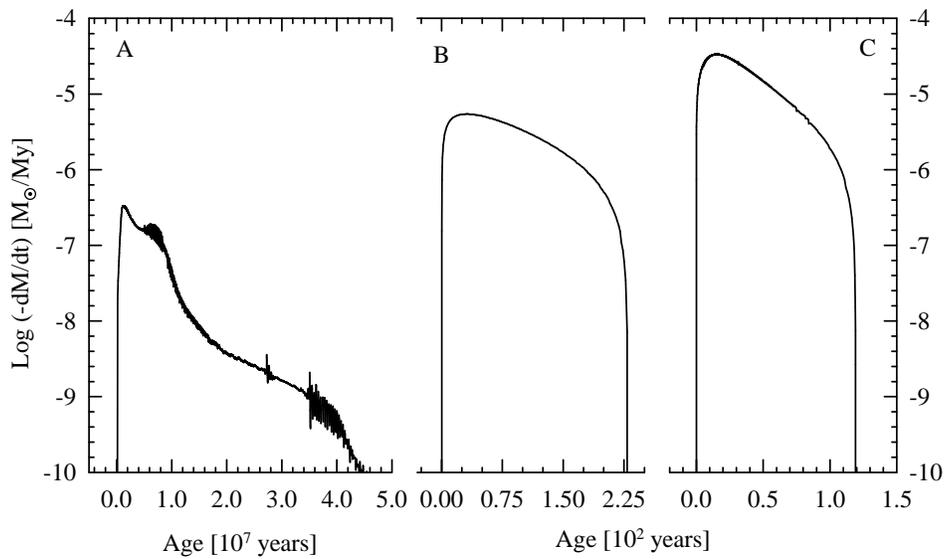}
\caption{The mass  transfer rate  from the primary  star of  system 1.
The  star undergoes  a  long  mass transfer  episode  soon after  core
hydrogen exhaustion  (panel A). In panels  B and C we  depict the mass
transfer  rate   corresponding  to   the  episodes  due   to  hydrogen
thermonuclear flashes. Notice that they  are, by far, shorter and with
a maximum rate about two to  three orders of magnitude larger than the
corresponding to  the first episode.   In each panel, time  is counted
from the onset of  mass transfer on. For the age of  the star at these
moments, see Table \ref{table_system_1}.}
\label{fig_mdot_1}
\end{figure*}

\noindent Here, $q_0  \equiv M  / M_{2}$ is  the initial  the mass ratio  of the
binary  components   and  $P_0$  the  initial   orbital  period.   The
evolutionary track for  the primary component of system  1 is shown in
Fig.   \ref{fig_HR_1}  and the  conditions  at  the beginning  (points
labelled with odd numbers) and end (points labelled with even numbers)
of mass transfer episodes are  included in Table \ref{table_system_1}. The star
fills its Roche  lobe soon after the end of core hydrogen burning.  
From then  on, the star begins to transfer mass  on a timescale of $\approx 10^7$ y (see Fig.  \ref{fig_mdot_1}).   The star ends
the  first mass transfer  episode with  a much  lower mass  value (see
Table \ref{table_system_1}) and begins to evolve bluewards
to  the white dwarf  stage. This evolution is stopped, because  the hydrogen  envelope is  still thick
enough to ignite  in a flash fashion, forcing the star  to inflate on a
very short  timescale. Then, the  star swells very suddenly  and fills
the corresponding  Roche lobe  again.  Then, the  star suffers  from a
second  mass  transfer  episode   depicted  in  the  second  panel  of
Fig. \ref{fig_mdot_1}. In this second episode the total amount of mass
transferred is  very tiny and occurs  on a very short  timescale. As a
consequence  the Roche lobe  size is  almost constant  and the  star is
forced to evolve along a constant - radius line. Now, after the end of
the  second  mass transfer  episode,  the  star  has a  thinner  outer
hydrogen layer  and evolves again  bluewards to the white  dwarf stage
but again it suffers from a second hydrogen thermonuclear flash. Then,
the history  is repeated: the star suddenly  swells up and  fills the
Roche lobe by third time. Mass transfer begins and again a very little
amount  of mass is  transferred but  now in  a much  shorter timescale
compared  to  the  case  of  the immediately  previous  mass  transfer
episode.   After ending  this third  mass transfer  episode,  the star
evolved  again bluewards  to  the  white dwarf  stage  and cools  down
without suffering any other flash.

As it  is well known,  hydrogen thermonuclear flashes occur  very near
the stellar surface.   As the hydrogen rich outer  layer is very thin,
if  we want  to  compute all  these  stages properly,  we  do need  to
consider a very  fine zoning for these critical  layers. As previously
mentioned,  at flash  - induced  mass transfer  episodes, a  very tiny
fraction of  the star is lost  before the star detaches  and starts to
evolve bluewards again.   Then, we do also need a  very thin zoning in
order to  account properly for the  exact amount of hydrogen  remaining in
the star, as  well as the large gravitational  energy release when the
star undergo  large radius  excursions. This is  the kind  of physical
situation that  suggested us the  convenience of designing  our binary
evolution code in the way  we have presented in the previous sections,
especially regarding the treatment of  the outer layers and the amount
of  matter we allow  to be  above the  outermost finite  - differenced
zone.

We remark that, assuming this  fully implicit scheme including the MTR
we have found that the most  difficult quantity to compute is just the
MTR.   This  is  especially  true  when the  outer  layers  become  in
convective equilibrium. This is  reflected in some strong oscillations
shown in the first panel of Fig. \ref{fig_mdot_1}. In any case we feel
this not to affect the  main course of the evolution. Remarkably, such
oscillations  are not  present in  the flash  - induced  mass transfer
episodes.

System 2 is a Class A system (see Fig. 4) which transfers
most of its mass when it  is still burning hydrogen in the core. Then,
it  suffers three  thermonuclear flashes  with its  corresponding mass
transfer episodes in a way very similar to the corresponding to System
1  (see Table \label{table_system_2}  and Figs. 5 and 6).

System 3 (Class B) suffers  from four mass transfer episodes, three of
them induced by each thermonuclear flash.  The main characteristics of
these evolutionary calculations are included in Figs. 7 - 9 and   in
Tables    \ref{table_system_2}   and
\ref{table_system_3} we include the conditions at the onset and end of
mass transfer episodes.

System  3 evolution  can be  compared  with the  results presented  by
Podsiadlowski et  al. (2001) in their  Fig. 14. They  evolved a system
composed initially by a 1.4 $M_{\odot}$ normal star and a neutron star
of the same  mass. Mass transfer begins near the  end of core hydrogen
burning.  In doing so,  they employed a full nuclear reaction network,
convective   overshooting,  and   different  assumptions   about  mass
transfer from the ones we have done here. In particular, they used $\beta= 0.5$,  i.e., that half of  the mass
lost by the normal star is accreted by the neutron star as long as the
rate   is  smaller   than   the  Eddington   limit  ($\approx   10^{-8}
M_{\odot}/y$). Also, they considered $\alpha= 1$ for the angular
momentum losses. Clearly, these are large differences
compared with our  System 3 and also with  the physical ingredients we
considered in the present paper.
Notice  that  Podsiadlowski  et   al.   (2001)  found  three  hydrogen
thermonuclear flashes,  two of which  leading to Roche lobe  overflow. In
our system, we  also have three hydrogen flashes,  all of them leading
to lobe overflow.  This is  an important difference that we suspect it
to be related with the  different assumptions in the orbital evolution
of the system. We  also consider that the differences in the
treatment of nuclear energy release are of key importance. Clearly, 
our equilibrium cycles predict  stronger flashes. However, notice that
the  mass of  the final  helium white  dwarf is  very similar  in both
calculation (0.199 $M_{\odot}$ for Podsiadlowski et al.  2001).

\begin{table}
\caption{Selected evolutionary stages of system 1}
\begin{tabular}{cccccc}
\hline Point  & $Age/10^{9} y$ & $M/M_{\odot}$  & $Log(L/L_{\odot})$ &
$Log(T_{eff})$  &  $P\; [d]$  \\  \hline 1  &  8.6344152  & 2.00000  &
1.470297 &  3.862974 & 1.500 \\ 2  & 9.0981136 & 0.23161  & 0.950006 &
3.833146 &  2.442 \\ 3 & 9.6587255  & 0.23161 & 2.026359  & 4.102230 &
2.442 \\ 4 & 9.6587278 & 0.23096  & 2.306213 & 4.171956 & 2.449 \\ 5 &
9.9650518 & 0.23096  & 1.740276 & 4.030472 & 2.449 \\  6 & 9.9650530 &
0.22935 & 2.608005 & 4.246795 & 2.468 \\ \hline
\end{tabular} \label{table_system_1}
\end{table}

\begin{table}
\caption{Selected evolutionary stages of system 2}
\begin{tabular}{cccccc}
\hline Point  & $Age/10^{9} y$ & $M/M_{\odot}$  & $Log(L/L_{\odot})$ &
$Log(T_{eff})$  & $P\;  [d]$  \\ \hline  1  & 0.42652401  & 2.00000  &
1.280174 & 3.925827  & 0.697 \\ 2 & 9.68186661 &  0.18421 & 0.179932 &
3.785182 & 1.006  \\ 3 & 10.1453763 & 0.18421 &  1.100802 & 4.015836 &
1.007 \\ 4 & 10.1453778 & 0.18395 & 1.163948 & 4.031070 & 1.008 \\ 5 &
10.1883408 & 0.18421 & 1.099444 & 4.015355 & 1.008 \\ 6 & 10.1883427 &
0.18348 &  1.341075 & 4.075097 & 1.011  \\ 7 & 10.2173678  & 0.18348 &
1.217356 & 4.044674  & 1.011 \\ 8 & 10.2173692 &  0.18313 & 1.417843 &
4.094100 & 1.013 \\ \hline
\end{tabular} \label{table_system_2}
\end{table}

\begin{table}
\caption{Selected evolutionary stages of system 3}
\begin{tabular}{cccccc}
\hline Point  & $Age/10^{9} y$ & $M/M_{\odot}$  & $Log(L/L_{\odot})$ &
$Log(T_{eff})$  &  $P\; [d]$  \\  \hline 1  &  2.0641016  & 1.40000  &
0.844531 &  3.797418 & 1.000 \\ 2  & 3.4962904 & 0.20567  & 0.472654 &
3.702205 &  2.803 \\ 3 & 3.8905910  & 0.20567 & 1.053372  & 3.847385 &
2.803 \\ 4 & 3.8905929 & 0.20530  & 1.278767 & 3.903560 & 2.809 \\ 5 &
3.9239877 & 0.20530  & 1.299815 & 3.908821 & 2.809 \\  6 & 3.9239887 &
0.20490 &  1.546744 & 3.970370  & 2.815 \\  7 & 4.0033926 &  0.20490 &
1.385939 &  3.930169 & 2.815 \\ 8  & 4.0033931 & 0.20415  & 1.853513 &
4.046706 & 2.827 \\ \hline
\end{tabular} \label{table_system_3}
\end{table}

Finally,  in Fig. 10 we  show the  evolution of  the central
part  of the  three  computed  objects in  the  temperature -  density
plane. As expected, the evolution  is rather similar and is limited to
temperatures  below helium  ignition. The  flash episodes  suffered by
these objects  are responsible for the oscillation  in the temperature
of the  star before it gets  near the final density  value.  Then, the
objects cool down almost at  constant central density, as it should be
expected.

These calculations indicate that the above described scheme is capable
of handling the  evolution of close binary systems  in a very adequate
way.   In our opinion, the here described code  should  be a
valuable tool  in handling binary  evolution.  Let us remark  that, we
have tested the code with low mass systems because are the ones we are
most interested in.  Also, we  have found very good convergence of the
method for the case of massive CBSs.

In closing, we  should remark that the above  described results should
not be considered  a state - of - the  - art evolutionary calculations
but a  serious test  of the  numerical scheme as  a previous  step for
future works.   This is so  mainly because of  the lack of  a detailed
nuclear reactions network  and also for the absence  of a treatment of
diffusion. Notice that diffusion is  a key process in the evolution of
objects  in the  range  of masses  we  have here  considered and  even
largely determine the total number of thermonuclear flashes the object
suffers from (see Althaus, et al. 2001).

%---------------------------------------------------------------------
\section{Discussion and Future Applications} \label{sec:discu}

In  this paper  we  have  presented a  Henyey-type  code tailored  for
computing stellar evolution in  close binary systems (CBSs). This code
is  a modification  of the  scheme presented  long ago  by Kippenhahn,
Weigert \& Hofmeister (1967) for the case of the evolution of isolated
objects.

Here we  considered that  mass transfer occurs  when  the star overflows its Roche lobe. The main  characteristic of the present  code is
that it automatically computes the  beginning and end of mass transfer
stages,  computing the MTR  of the  object in  a fully  implicit, self
consistent fashion.   Perhaps, main shortcoming of the  scheme we have
presented is that it is not able to compute common envelope evolution,
simply because at these stages it is not possible to equate the radius
of the star to that of the Roche lobe.

In the present version of  the code, we have considered radiative OPAL
and conductive opacities,  neutrino emissivities, convection following
the mixing length prescription, and a detailed equation of state.  The
nuclear  reactions   have  been  considered  only  by   means  of  the
equilibrium cycles.   Regarding the physics of mass  transfer, we have
considered  the possibility that  a fraction  of the  mass transferred
from the primary is accreted  by the secondary.  Also we have included
angular  momentum  dissipation  due  to  gravitational  radiation  and
magnetic braking.

We have found that the scheme detailed above is able to handle most of
the  evolutionary stages  suffered  by  a donor  star  belonging to  a
CBS. Apart from the results  we have presented as a first application,
we have also tested  it for the case of massive CBSs  finding it to be
very adequate also for these purposes. Generally speaking, the code is
able to  perform single stellar evolution  and also to  find the onset
and end of each mass transfer episode. In all the test calculations we
have performed we  found it much more easy to  compute a mass transfer
rates (MTRs) when  outer layers are in radiative  equilibrium.  In the
case of convective layers convergence  is attained but sometimes it is
necessary a larger number of iterations.

As  a  first application  of  this  code  to astrophysically  relevant
conditions, we  have applied it  to the case  of the formation  of low
mass helium white dwarfs. These  computations should not be considered
as a state - of - the - art ones but as a test of the numerical scheme
as a previous step for future  works. This is so mainly because of the
lack of a detailed nuclear  reactions network and also for the absence
of a treatment  of diffusion.  

We expect that the physical and numerical performance of the code should be
largely improved  if we  replace the outer  boundary condition  for the
stellar  radius by  the formulation  of  Ritter (1988)  for mass  loss
transfer (see  the Introduction). This should  be especially important
at the beginning  and end of each mass transfer  episode and/or in the
case of  very low mass transfer  rates. Also, the  implementation of a
moving  grid should  help  in  producing smooth  curves  of mass  loss
vs. time.

As a  future application  for the  present code we plan to  study the
problem of the formation of helium white dwarfs including the complete
physical ingredients we  have considered in the previous  works of our
group on this topic. Also, we plan to study the relevance of the
irradiation  (Tout, et  al.   1989;  D'Antona \&  Ergma  1993) in  the
evolution of  the members of  CBSs. This should be  especialy relevant
for the case  of LMXBs which in some  evolutionary stages are expected
to undergo mass transfer driven by irradiation (Podsiadlowski 1991).

Fig. 4 - 10 are available upon request to the authors at their e-mail 
addresses.

%---------------------------------------------------------------------
\section{Acknowledgements}

The  authors want  to warmly  acknowledge our  referee  Prof.  Phillip
Podsiadlowski  for his very constructive  criticism and  advice. This  has
enabled  us not  only to  improve the  original version  of  this work
significantly but also has helped us on how to upgrade the performance
of the code  we presented here from a physical  and numerical point of
view.

Also, OGB wants to acknowledge Prof. Francesca D'Antona for discussion
about the problem of computing binary evolution.

%---------------------------------------------------------------------

\newpage
%----------------------------------------------------------------
%---               FIGURE               CAPTIONS              ---
%----------------------------------------------------------------

\begin{figure*}
\epsfysize=550pt \epsfbox{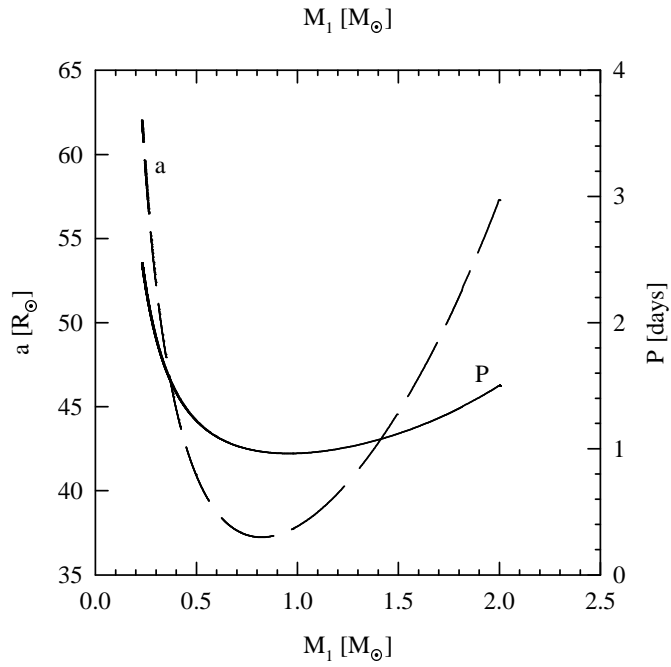}
\caption{The Orbital period and semiaxis  as a function of the mass of
the primary star of system 1.}
\label{fig_orbita_1}
\end{figure*}

\bsp

\label{lastpage}

\end{document}